\documentclass[10pt,journal,compsoc]{IEEEtran}
\usepackage{url}
\usepackage{multirow}
\usepackage[english]{babel}
\usepackage{float}
\usepackage{breqn}
\usepackage{footnote}
\usepackage{balance}
\usepackage{blindtext, graphicx}
\usepackage{booktabs}
\usepackage{soul,color}
\usepackage{algorithm}
\usepackage{bm}
\usepackage{ifpdf}
\usepackage{amsmath}
\usepackage{array}
\usepackage{amssymb}
\usepackage{amsfonts}
\usepackage{textcomp}
\usepackage{xcolor}
\usepackage{tikz}
\usepackage{colortbl}
\usepackage[noend]{algpseudocode}
\usepackage{enumitem}
\usepackage{fancyhdr}
\newcommand{\cmmnt}[1]{}  
\newcommand{\mytb}{\cellcolor{gray!25}}

\newcommand\encircle[1]{%
\tikz[baseline=(X.base)] 
  \node (X) [draw, scale=0.75, shape=circle, inner sep=0, fill=black, text=white, minimum size=0em] {\strut #1};}

\usepackage{caption}

\usepackage{tikz}

\usepackage{pifont}
\usepackage[noadjust]{cite}

\ifCLASSINFOpdf

\else

\fi
\usepackage[capitalise]{cleveref}

\begin{document}

\title{A Near-Sensor Processing Accelerator for\\ Approximate Local Binary Pattern Networks}

\author{Shaahin Angizi,
Mehrdad Morsali,
Sepehr Tabrizchi, and
Arman Roohi \vspace{-0.5em} 
         \thanks{This work is supported in part by the National Science Foundation under Grant No. 2228028, 2216772, and 2216773.}
\IEEEcompsocitemizethanks{\IEEEcompsocthanksitem S. Angizi and M. Morsali are with the Department of Electrical and Computer Engineering, New Jersey Institute of Technology, Newark, NJ, USA.
E-mail: shaahin.angizi@njit.edu.
\IEEEcompsocthanksitem A. Roohi and S. Tabrizchi are with School of Computing, University of Nebraska–Lincoln, Lincoln NE, USA.
E-mail: aroohi@unl.edu.}
}

\markboth{}%
 {Shell \MakeLowercase{\textit{et al.}}: Bare Demo of IEEEtran.cls for IEEE Transactions on Magnetics Journals}

\IEEEtitleabstractindextext{%
\begin{abstract}
In this work, a high-speed and energy-efficient comparator-based \underline{N}ear-\underline{S}ensor \underline{L}ocal \underline{B}inary \underline{P}attern accelerator architecture (NS-LBP) is proposed to execute a novel local binary pattern deep neural network. First, inspired by recent LBP networks, we design an approximate, hardware-oriented, and multiply-accumulate (MAC)-free network named Ap-LBP for efficient feature extraction, further reducing the computation complexity. Then, we develop NS-LBP as a processing-in-SRAM unit and a parallel in-memory LBP algorithm to process images near the sensor in a cache, remarkably reducing the power consumption of data transmission to an off-chip processor. Our circuit-to-application co-simulation results on MNIST and SVHN data-sets demonstrate minor accuracy degradation compared to baseline CNN and LBP-network models, while NS-LBP achieves 1.25 GHz and energy-efficiency of 37.4 TOPS/W.
NS-LBP reduces energy consumption by 2.2$\times$ and execution time by a factor of 4$\times$ compared to the best recent LBP-based networks.
\end{abstract}

\begin{IEEEkeywords}
Processing-in-memory, accelerator, near-sensor processing, SRAM. \vspace{-0.2em}
\end{IEEEkeywords}}\vspace{-1em}

\maketitle

\IEEEdisplaynontitleabstractindextext

\IEEEpeerreviewmaketitle

\vspace{-4.2em}
\section{Introduction}
\IEEEPARstart{I}{nternet} of things' (IoT) nodes consist of sensory systems, which enable massive data collection from the environment and people to process with on-/off-chip processors ($10^{18}$ ops). In most cases, large portions of the captured sensory data are redundant and unstructured. Data conversion and transmission of large raw data to a back-end processor imposes high energy consumption, high latency, and low-speed feature extraction on the edge \cite{hsu2019ai}.  To overcome these issues, computing architectures will need to shift from a cloud-centric approach to a thing-centric (data-centric) approach, where the IoT node processes the sensed data. 
This paves the way for a new smart sensor processing architecture \cite{yamazaki20174,hsu20200}, in which the pixel's digital output is accelerated near the sensor leveraging an on-chip processor. Unless a Processing-in-Memory (PIM) mechanism is exploited \cite{li2017drisa,eckert2018neural} in this method, the von-Neumann computing model with separate memory and processing blocks connecting via buses imposes long memory access latency, limited memory bandwidth, and energy-hungry data transfer restricting the edge device's efficiency and working hours \cite{hsu2019ai,song2017pipelayer}. The main idea of PIM is to incorporate logic computation within memory units to process data internally. 

From the computation perspective, numerous artificial intelligence applications require intensive multiply-accumulate (MAC) operations, which contribute to over 90\% of various deep Convolutional Neural Networks (CNN) operations \cite{eckert2018neural}.
Various processing-in-SRAM platforms have been developed in recent literature \cite{eckert2018neural,aga2017compute,liu2020ns,jiang2020c3sram,yin2020xnor}. Compute cache \cite{aga2017compute} supports simple bit-parallel operations (logical and copy) that do not require interaction between bit-lines.
Neural Cache \cite{eckert2018neural} presents an 8T transposable SRAM bit-cell and supports bit-serial in-cache MAC operation. Nevertheless, this design imposes a very slow clock frequency and a large cell and Sense Amplifier (SA) area overhead.  In \cite{lee2020bit}, a new approach to improve the performance of the Neural Cache has been presented based on 6T SRAM, enabling faster multiplication and addition with a large SA overhead.
While the presented designs show acceptable performance over various image data-sets by reducing the number of operations, i.e., MACs, using shallower models, quantization, pruning, etc., they are essentially developed to execute the existing CNN algorithms that lead to a gap between meets and needs. 
We believe such a discrepancy can be avoided by \emph{co-developing an intrinsically-low computation network and an efficient PIM platform on the sensor side}. 
Regarding the model reduction of CNNs, Local Binary Pattern (LBP)-based implementations have attained worldwide attention for edge devices, resulting in a similar output inference accuracy \cite{juefei2017local,sarwar2017gabor,lin2020local}. More interestingly, the amount of convolution operations is drastically reduced owing to the sparsity of kernels and conversion to simpler operations such as addition/subtraction \cite{LBCNN} and comparison \cite{angizi2018cmp}.

In this work, inspired by recent LBP networks, (1) we first develop a novel approximate, hardware-oriented, and MAC-free neural network named Ap-LBP in Section 3 to reduce computation complexity and memory access by disregarding the least significant pixels to perform efficient feature extraction. The Ap-LBP is leveraged on the sensor side to simplify LBP layers before even mapping the data into a near-sensor memory; (2) NS-LBP is designed as a comparator-based processing-in-SRAM architecture, in conjunction with the LBP parallel in-memory algorithm in Section 4, which remarkably reduce the power consumption as well as the latency of data transmission to a back-end processor; (3) In Section 5, we propose a correlated data partitioning and hardware mapping methodology to process the network locally; and (4) We extensively evaluate NS-LBP performance, energy efficiency, and inference accuracy trade-off compared to recent designs with a bottom-up evaluation framework in Section 6.\vspace{-0.7em}

\section{Background \& Motivation}
\subsection{Near-Sensor \& In-Sensor Processing} 
Systematic integration of computing and sensor arrays has been widely studied to eliminate off-chip data transmission and reduce Analog-to-Digital Converters (ADC) bandwidth by combining CMOS image sensor and processors in one chip as known as Processing-Near-Sensor (PNS) \cite{hsu20200,bhowmik2019event,angizi2019mrima,yamazaki20174,bong2017low,bhowmik2019visual,agrawal2018x,Angizi2020Assem,angizi2019acceleratinga}, or even integrating pixels and computation unit so-called Processing-In-Sensor (PIS) \cite{xu2020macsen,park20147,angizi2022pisa,xu2021senputing,li20215,xu2020utilizing,xu20214,angizijxcdc,tabrizchi2022design}.
However, since enhancing the throughput and increasing the computation load on the resource-limited IoT devices is followed by a growth in the temperature and power consumption as well as noises that lead to accuracy degradation \cite{chu2014neuromorphic,xu2020macsen}, the computational capabilities of PNS/PIS platforms have been limited to less complex applications \cite{hsu2019ai,bong201714,mukherjee2019fast,angizi2019graphide}.
This includes particular feature extraction tasks, e.g., Haar-like image filtering \cite{bong201714} and blurring \cite{hsu20200}.
Various powerful processing-in-SRAM (in-cache computing) accelerators have been developed in recent literature that can be employed as a PNS unit \cite{liu2020ns,jiang2020c3sram,yin2020xnor,aga2017compute,eckert2018neural,yin2019vesti,wang201928,simon2019fast,yang201924,lee2020bit}. Compute cache \cite{aga2017compute} supports simple bit-parallel operations (logical and copy) that do not require interaction between bit-lines. XNOR-SRAM \cite{yin2020xnor} accelerates ternary-XNOR-and-accumulate operations in binary/ternary Deep Neural Networks (DNNs) without row-by-row data access. C3SRAM \cite{jiang2020c3sram} leverages  capacitive-coupling computing to perform XNOR-and-accumulate operations for binary DNNs. However, both XNOR-SRAM and C3SRAM impose huge overhead over the traditional SRAM array by directly modifying the bit-cells.
Neural Cache \cite{eckert2018neural} presents an 8T transposable SRAM bit-cell and supports bit-serial in-cache MAC operation. Nevertheless, this design imposes a very slow clock frequency and a large cell and SA area overhead.  In \cite{lee2020bit}, a new approach to improve the performance of the Neural Cache has been presented based on 6T SRAM, enabling faster multiplication and addition with a large SA overhead. In the PIS domain, a CMOS image sensor with dual-mode delta-sigma ADCs is designed in \cite{kim2020chip} to process  1$^{st}$-convolutional layer of Binarized-Weight Neural Networks (BWNNs). RedEye \cite{redeye} executes the convolution operation using charge-sharing tunable capacitors. Although this design shows energy reduction compared to a CPU/GPU by sacrificing accuracy, to achieve high accuracy computation, the required energy per frame increases dramatically by 100$\times$. 
Macsen \cite{xu2020macsen} processes the 1$^{st}$-convolutional layer of BWNNs with the correlated double sampling procedure achieving 1000fps speed in computation mode. However, it suffers from humongous area-overhead and power consumption.
There are \textbf{three main bottlenecks} in IoT imaging systems that this paper aims to solve: {(1) The data access and movement consume most of the power ($>90\%$ \cite{xu2020macsen,yin2019vesti,choi2015energy}) in conventional image sensors; (2) the computation imposes a large area-overhead and power consumption in more recent PNS/PIS units and requires extra memory for intermediate data storage; and (3) the system is hardwired so their performance is intrinsically limited to one specific type of algorithm or application domain, which means that such accelerators cannot keep pace with rapidly evolving software algorithms.}  \vspace{-1.2em} 

\subsection{LBP-based Networks}
\begin{figure}[t]
\includegraphics [width=\linewidth]{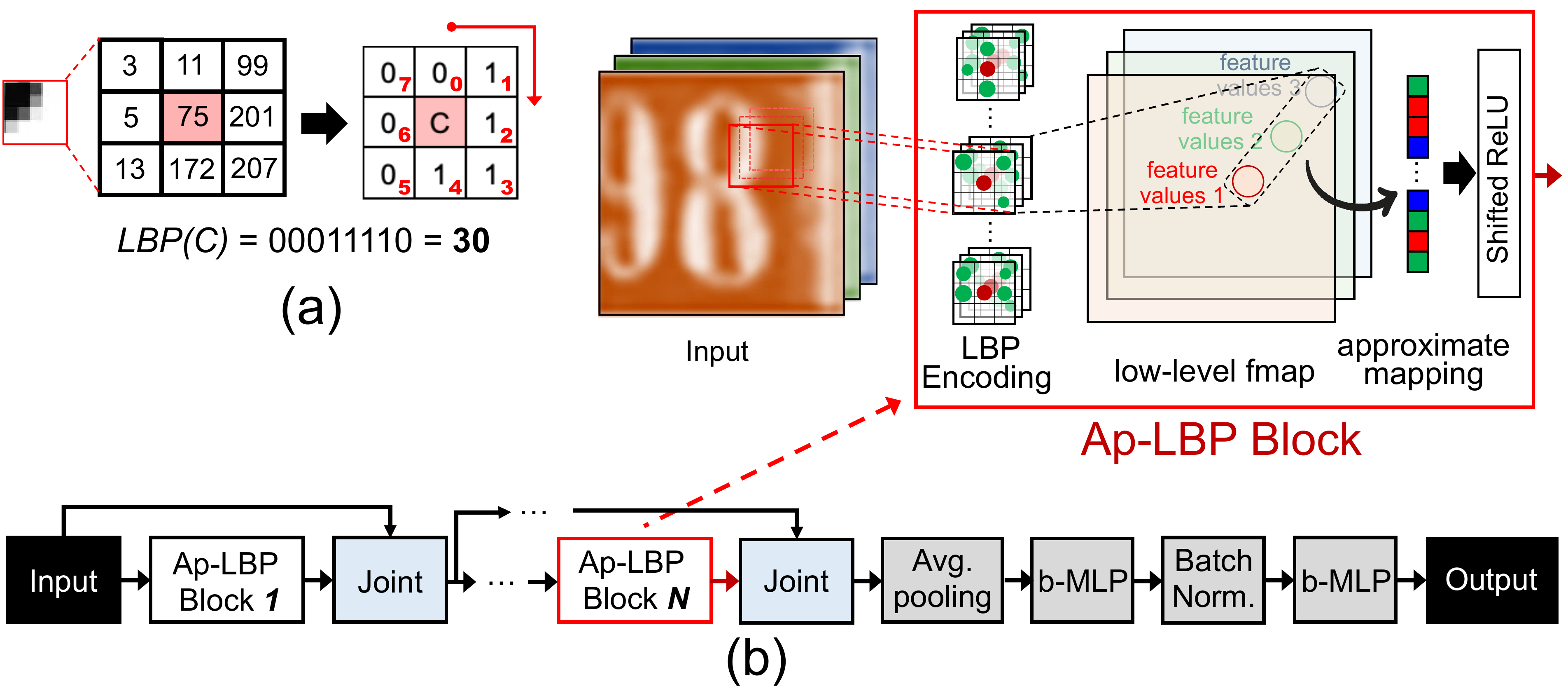}\vspace{-2em}\\
\caption{(a) Standard LBP encoding with $3\times 3$ descriptor size, (b) The structure of the Ap-LBP, with \emph{N} Ap-LBP blocks.}
\label{lbp}\vspace{-1em}
\end{figure}

An LBP kernel is a computationally efficient feature descriptor that scans through the entire image like that of a convolutional layer in a CNN. 
The LBP descriptor is formed by comparing the intensity of surrounding pixels serially with the central pixel, referred to as Pivot, in the selected image patch. Neighbors with higher (/lower) intensities are assigned with a binary value of `1'(/`0') and finally the bit stream is sequentially read and mapped to a decimal number as the feature value assigned to the central pixel, as shown in Fig. \ref{lbp}(a). The LBP encoding operation of central pixel $C(x_{c},y_{c})$ and its reformulated expression can be mathematically described as $LBP(C)=\sum_{n=0}^{d-2}cmp(i_{n},i_{c}) \times 2^{n}$ \cite{juefei2017local}, where $d$ is the dimension of the LBP, $i_{n}$ and $i_{c}$ represent the intensity of $n\textsuperscript{th}$ neighboring- and central-pixel, respectively; thus, $cmp(i_{n},i_{c})=1$ when $i_{n}\geq i_{c}$, otherwise outputs 0. Simulating LBP is accomplished using a ReLU layer and the difference between pixel values. As part of the training process, (i) the sparse binarized difference filters are fixed, (ii) the one-by-one convolution kernels function as a channel fusion mechanism, and (iii) the parameters in the batch normalization layer (batch norm) are learned. 

\begin{figure}[t]
\centering
\includegraphics [width=0.9\linewidth, height= 2.5cm]{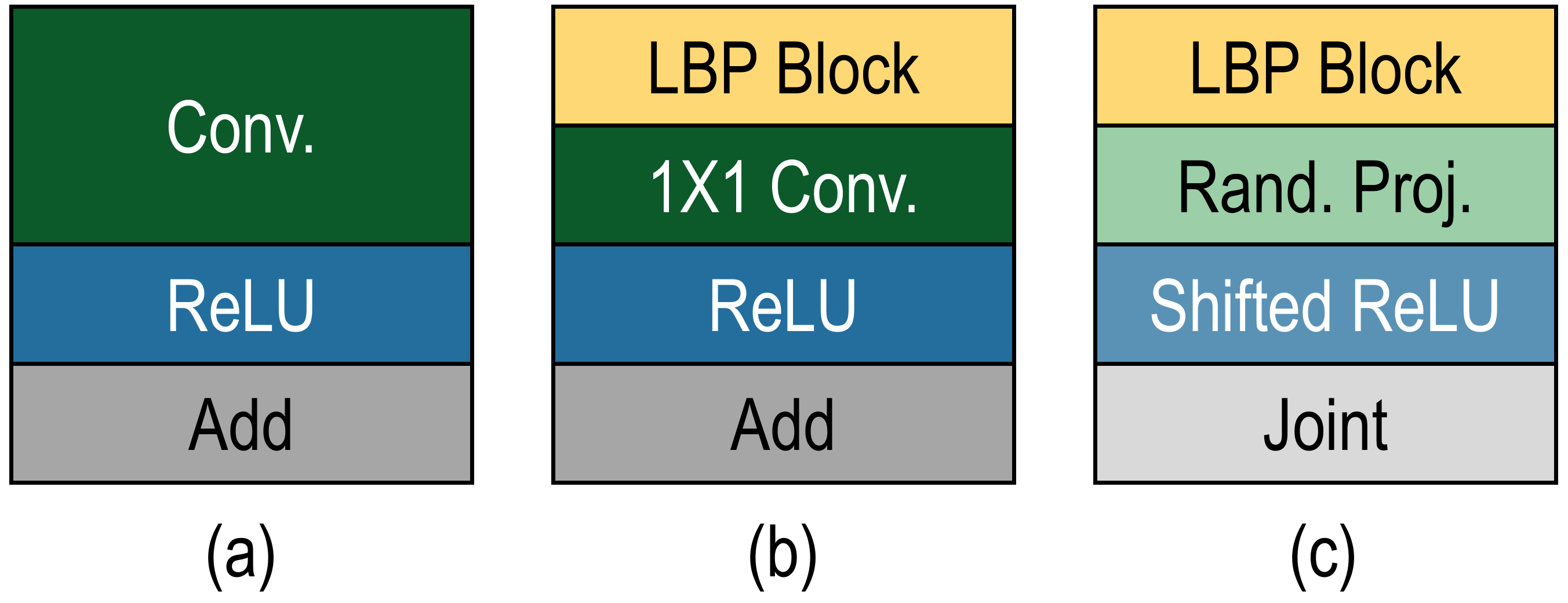}\\ \vspace{-1em}
\caption{Different basic building blocks of a (a) residual network, (b) LBCNN \cite{LBCNN}, and (c) LBPNet \cite{LBPNET}.}\vspace{-1em}
\label{lbps}
\end{figure}

The Local Binary Pattern Network (LBPNet) \cite{LBPNET} and Local Binary Convolutional Neural Network (LBCNN) \cite{LBCNN} are two recent LBP networks where the convolutions are approximated by local binary additions/subtractions and local binary comparisons, respectively. It should be noted that LBPNet and LBCNN are quite different, despite their similarity in their names, as illustrated in Fig. \ref{lbps}. 
In the LBCNN, batch norm layers are still heavily utilized, which are completed in floating-point numbers for the linear transform. Moreover, since the size and computation of 2D batch norm layers are linear in the size of the feature maps, model complexity increases dramatically. Therefore, the use of LBCNNs for resource-constrained edge devices, such as sensors, is still challenging and impractical.
LBPNets, on the other hand, learn directly about the sparse and discrete LBP kernels, which are typically as small as a few KBs. By using LBPNet, the computation of dot products and sliding windows for convolution can be avoided. Rather, the input is sampled, compared, and then the results of the comparisons are stored in determined locations. Therefore, in LBPNet, only trained patterns of sampling locations are held, and no MAC operations (convolution-free) are performed, making it a hardware-friendly and suitable model for edge computing.

\section{A\MakeLowercase{p}-LBP Network}\vspace{-0.2em}
\label{LBP}

\begin{figure}[t]
\begin{center}
\includegraphics [width=0.95\linewidth, height=6.4cm]{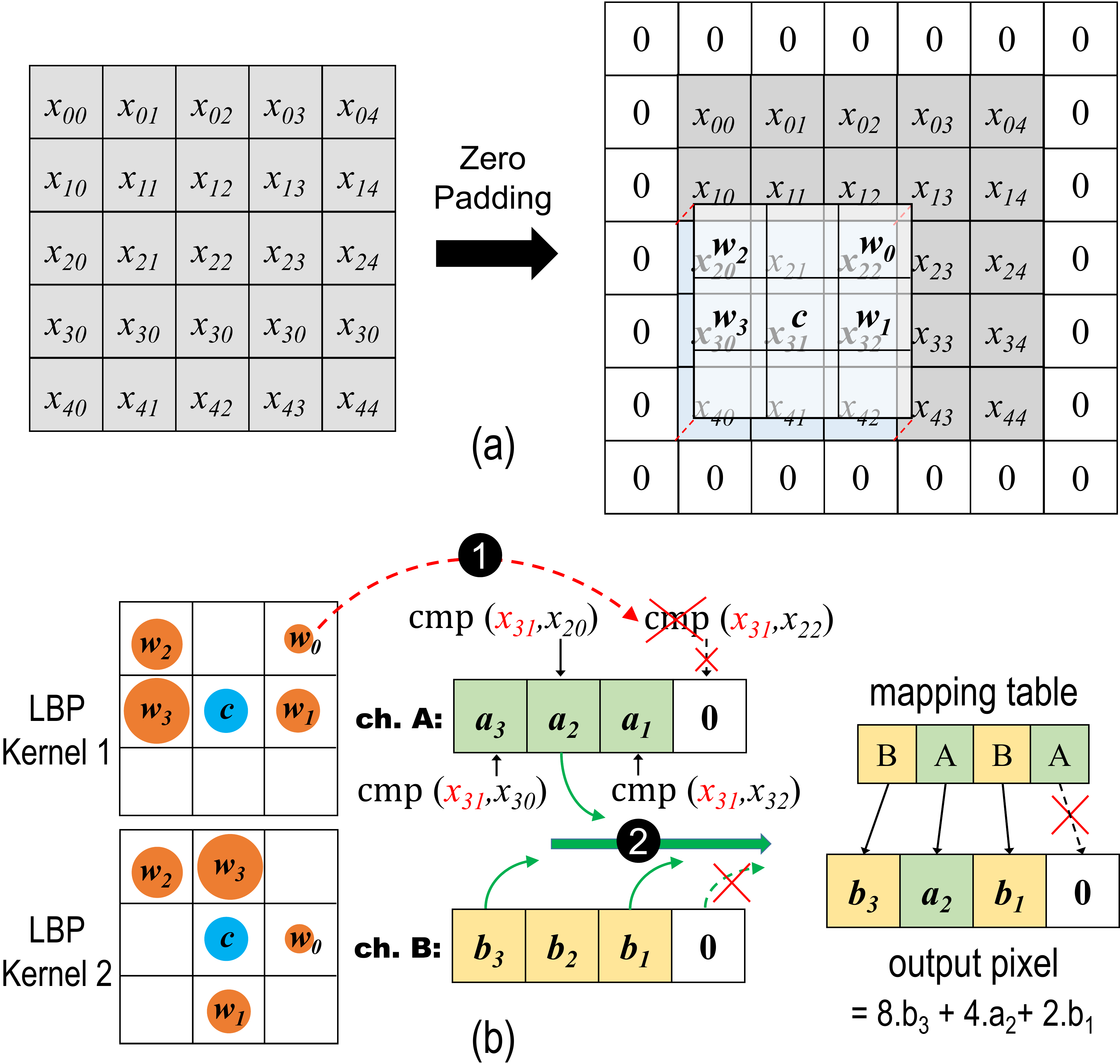} \vspace{-1.0em}\\
\caption{(a) Visualization of $X$ and $X^{p0}$, zero padding; the blue boxes is the sliding $3\times 3$ LBP kernel, (b) Approximate mapping using comparison and memory access skipping schemes on channels \textbf{A} and \textbf{B} to generate ofmap.}\vspace{-1em}
\label{pac}
\end{center}
\end{figure}

The Ap-LBP network is trained similarly to the LBPNet, which learns a set of local binary patterns. The Ap-LBP structure, visualized in Fig. \ref{lbp}(b), consists of multiple LBP layers followed by an average pooling, two Multi-Layer Perceptron blocks (MLP), and one batch normalization layer. A standard convolutional layer is replaced with a layer using LBPs, which means neither multiplication nor addition is required, and MAC operations are performed via memory access and comparison. An LBP layer, including an LBP Block and a Joint operation, is leveraged to extract feature maps. Each LBP block consists of an LBP Encoding step that can be readily implemented by a comparator\footnote{ In the backward propagation, binary comparisons are replaced by a modified hyperbolic tangent (tanh) function and shifted to become differentiable.} to generate new feature maps connected to an approximate mapping and shifted-ReLU blocks to increase nonlinearity. The output of the LBP blocks is cascaded with the input feature maps (ifmaps) using joint blocks.
Figure \ref{pac} illustrates a portion of the LBP block's operation. In the Ap-LBP, the size of the output feature maps (ofmaps) remains identical to the size of the ifmaps. To do so, the zero-padding approach might be utilized and the degree of zero insertions is calculated by $pad=[s\times (out-1)-in+f]/2$, where {\emph{s}} is the stride window's size, {\emph{out}} and {\emph{in}} are sizes of the ofmap and the ifmap, respectively, and {\emph{f}} is size of the LBP kernel. This expression works perfectly for square matrices.
For example, as shown in Fig. \ref{pac}(a) with \emph{s}=1, \emph{in}= 5 and \emph{f}= 3, to produce an ofmap with \emph{out}= 5, zero-padding approach with a degree of one should be utilized.

The learned sets of LBPs from the training step are used in the encoding part to denote the sampling points in ifmaps' positions that are to be compared with a pivot. After the training phase, pre-defined locations in encoding matrices and bit arrays are determined and remain fixed during the inference phase, e.g., LBP Kernels 1 and 2 in Fig. \ref{pac}(b). 
Since the given weights to the thresholded (compared) values and mapping patterns are specified, a \textit{Partial Approximate Computing method (PAC)} is developed to further improve the performance at the cost of lower accuracy.
The PAC includes two primary operations: (1) Skip comparison: since in the LBP layer, the positions (weights) of LBP kernels' elements are already specified, pixel-to-pivot comparison operation related to the Least Significant Bits (LSB) can be omitted, and ofmap is written by zero, step \encircle{1} shown in Fig. \ref{pac}(b). (2) Skip memory access: in the LBP channel fusion step, a pre-defined mapping table, referred to as a projection map, is fixed for all outputs within the same channel to generate an output pixel passing through the shifted-ReLU function. Accordingly, read (/write) operations from (/to) the LSBs of the channel's responses (output pixels) can be skipped, as shown in step \encircle{2} shown in Fig. \ref{pac}(b). 
By leveraging the PAC, comparison operations and memory accesses can be reduced as significant portions of Ap-LBP computation in order to minimize energy consumption with minimum accuracy loss. For example, in Fig \ref{pac}(b), the original LBPNet implementation requires 8 comparisons, 14 read and 12 write operations; however, using Ap-LBP, the output pixel can be generated by 6, 11, and 9 comparisons, read and write operations, respectively, which shows a considerable enhancement. The total number of operations required to produce output pixels in the LBP blocks, utilizing LBPNet and Ap-LBP can be computed by the following expressions:
\begin{equation}
\footnotesize
       OP_{LBPNet} = \underbrace{[e\times ch + m]}_{\#read} + \underbrace{[(e-1)\times ch]}_{\#comparison} + \underbrace{[(e-1)\times ch + m]}_{\#write}\\
\end{equation}
\begin{equation}
\footnotesize
    \begin{split}
        OP_{Ap\_LBP} & = \underbrace{[(e-apx)\times ch + m - apx]}_{\#read} + \underbrace{[(e-apx-1)\times ch]}_{\#comparison}\\ 
        & + \underbrace{[(e-apx-1)\times ch + m - apx]}_{\#write} \vspace{-2em}
    \end{split}\vspace{-1em}
\end{equation}
where {\emph{e}} is the number of LBP kernels' elements (number of samplings), {\emph{ch}} is the number of channels, {\emph{m}} is the number of mapping tables' elements, and {\emph{apx}} is the number of approximated bits. Increasing \emph{apx} results in higher speed and energy efficiency at the cost of accuracy degradation.

\begin{figure}[t]
\begin{center}
\includegraphics [width=0.85\linewidth]{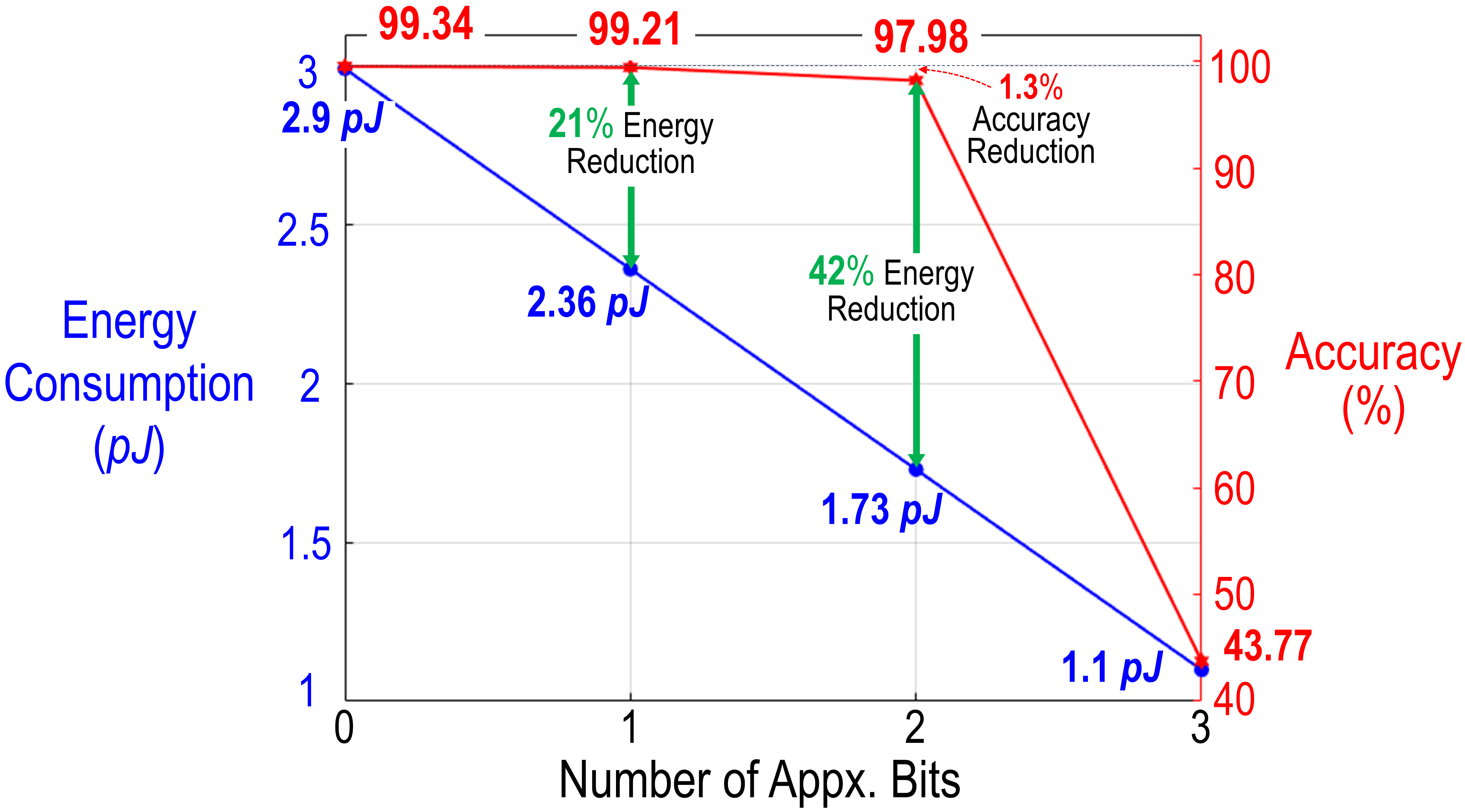}\\
\caption{Energy consumption vs. accuracy regarding the number of approximated bits on MNIST data-set.}
\label{acc} \vspace{-2em}
\end{center}
\end{figure}

\begin{figure*}[t]
\begin{center}
\begin{tabular}{c}
\includegraphics [width=0.97\linewidth]{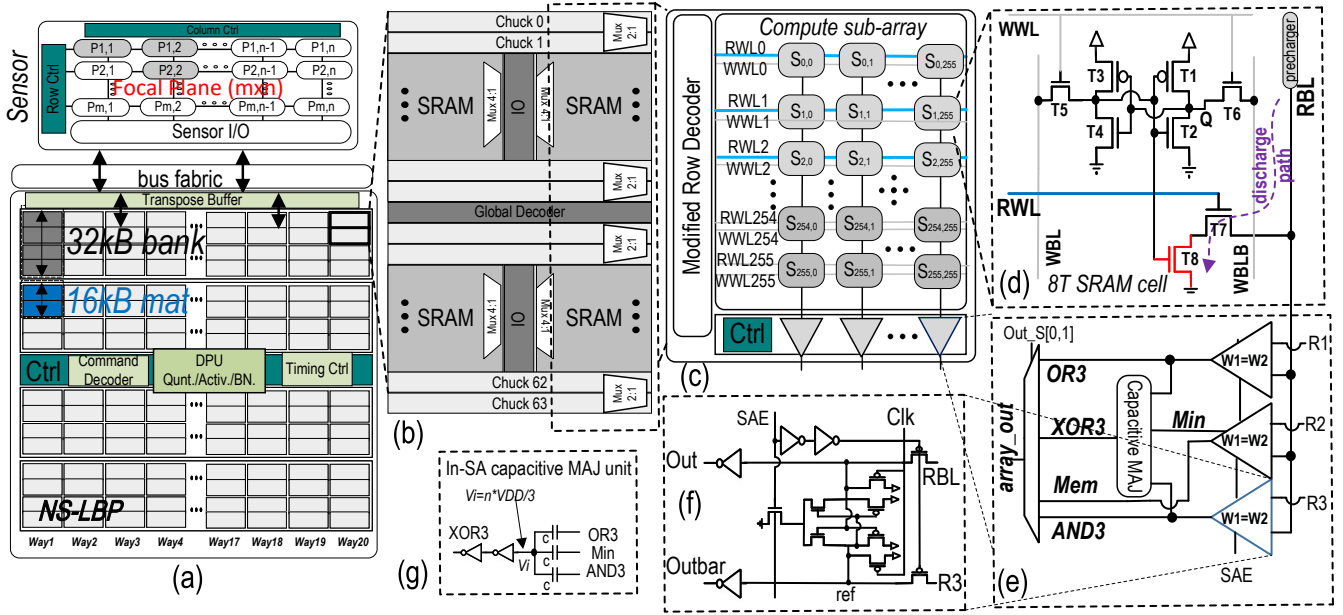} \vspace{-0.8em}\\
 \end{tabular} 
\caption{(a) The NS-LBP's geometry of a single 2.5MB cache slice connected to a sensor, (b) Computational matrix, (c) An 8KB SRAM computational sub-array, (d) 8T SRAM cell, (e) Proposed single-cycle SA design for LBP extraction, (f) Inside SA unit, (g) Capacitive majority function generator circuit.} \vspace{-1.5em}
\label{arc}
\end{center}
\end{figure*}
Figure \ref{acc} illustrates the accuracy results for Ap-LBP on MNIST with respect to the number of approximated bits and energy consumption of the LBP layers. The optimal condition occurs when 2 of 4 mapping tables' bits are approximated, which leads to relatively high energy savings (42\%) despite a small reduction in accuracy (1.3\%).
In the convolutional layer, the dimension of filters (ifmaps) is 4-D, $K\times ch\times r\times s$ ($M\times ch\times h\times w$), where $K$ and $M$ are the number of filters and ifmap, respectively, $ch$ is the number of channels, $r\times s$ is the spatial dimension of filters and $h\times w$ is the dimension of 2-D ifmaps. So the generated ofmaps' dimensions are $M\times K\times p\times q$, where $p\times q$ is ofmap's 2-D dimensions. The computational and memory costs for the convolution layer of both conventional CNNs and Ap-LBP networks are presented in Table \ref{table_cost}. To simplify matters, a single kernel ($K=1$) and a single ifmap ($M=1$) are considered. Since the difference between the number of samplings in an LBP pattern, $e$, and the number of approximated bits, $apx$, is relatively smaller than the spatial dimensions of kernels, that Ap-LBP, with MAC-free LBP layers, significantly reduces the hardware cost, both computation, and memory.
\begin{table}[h]
\centering
\caption{Hardware cost analysis of CNN vs. Ap-LBP.}\vspace{-0.6em}
\label{table_cost}
\scalebox{0.79}{
\begin{tabular}{ c c c c }
\hline
\mytb & \multicolumn{2}{c}{\mytb Computational cost} & \multirow{2}{*}{\mytb \begin{tabular}[c]{@{}c@{}}Memory \\ cost\end{tabular}} \\ \cline{2-3}
\multirow{-2}{*}{\mytb Network} & \mytb Mul--$O(N^2)$ & \mytb Add/Sub/Cmp--$O(N)$ & \mytb cost\\ \hline
CNN & $p\cdot q\cdot ch\cdot r\cdot s$ & $p\cdot q\cdot ch\cdot r\cdot s$ & $p\cdot q\cdot r\cdot s$ \\ \hline
\textbf{Ap-LBP} & - & $ch\cdot p\cdot q\cdot (e-apx)$ & $p\cdot q\cdot (e-apx)+(m-apx)$\\ \hline
\textbf{$\dfrac{Ap-LBP}{CNN}$} & 0 & $\dfrac{(e-apx)}{r\cdot s}$ &  $\dfrac{(e-apx)}{r\cdot s}+\dfrac{(m-apx)}{p\cdot q\cdot r\cdot s}$\\ \hline
\end{tabular}\vspace{-2.5em}
}
\end{table}

\vspace{-1.5em}

\section{Proposed NS-LBP}\vspace{-0.2em}
\subsection{Architecture}
We propose NS-LBP as a cache-based near-sensor architecture to accelerate the Ap-LBP network with a parallel in-memory LBP algorithm. NS-LBP's geometry of a single 2.5MB cache connected to an image sensor is shown in Fig. \ref{arc}(a).
A rolling-shutter CMOS image sensor is composed of m$\times$n photodiode-based pixels, which utilize the Correlated Double Sampling (CDS) mechanism \cite{xu2020macsen}. CDS measures the photodiode's voltage drop before and after an image light exposure and utilizes an ADC to convert it to a digital value. However, a significant amount of power is consumed by ADC conversion of raw images and high-throughput transmission \cite{hsu2019ai,xu2020macsen,hsu20200}.
To reduce the power consumption imposed by ADC and data transmission to the memory, we first modify the sensor controller and peripheral circuitry so that Ap-LBP's approximation can be applied on the sensor side by simply avoiding pixel conversion for less significant bits. This is explained in Section 3. By using this mechanism, the NS-LBP is assured of receiving only compute pixels and pivots. Cache slices within NS-LBPs are designed to have 80-32KB memory banks organized in 20 distinct ways. Each bank contains two 16KB memory matrices-mat (see Fig. \ref{arc}(a)). The centralized control unit (Ctrl) manages the internal memory data transfer, intra-bank computation, and a digital processing unit (DPU) common to all memory banks. The main computational cores of NS-LBP are 8KB computational sub-arrays as depicted in Fig. \ref{arc}(b)-(c).

According to our observations of existing sub-array-level processing-in-SRAM platforms, they face various challenges, such as multi-cycle in-memory operations, word-line underdrives, high-latency, read disturbances, etc. \cite{lee2020bit,wang201928,yang201924}, when it comes to comparison and addition operations required by the proposed Ap-LBP.
The proposed NS-LBP's sub-array (Fig. \ref{arc}(c)) leverages the voltage discharging profile of the read-write-decoupled 8T SRAM cell (Fig. \ref{arc}(d)) on Read-BL (RBL) used for the standard read operation and elevates it to implement Boolean logic between operands located in different memory rows in a single SRAM read cycle. 
In this way, we develop a processing-in-SRAM sub-array through a three-row activation mechanism by modifying the memory row decoder, SA, and Ctrl. 
It is important to note that the key idea comes from the observation that certain discharge rates on the precharged RBL can be expected based on selected memory bits. For instance, by activating three memory rows via Read Word-Lines (RWL), e.g., RWL0-RWL2 shown in Fig. \ref{arc}(c), if $S_{0,0}$, $S_{1,0}$, and $S_{2,0}$ memory cells hold binary ``1'', then the read access transistors (T8 in Fig. \ref{arc}(d)) remain OFF, and the RBL precharged voltage doesn't degrade. However, if all cells hold binary ``0'', the RBL voltage is rapidly discharged through T8s. Accordingly, we propose a new reconfigurable SA as shown in Fig. \ref{arc}(e) consisting of three sub-SAs, each dedicated to computing a particular function. 
With a proper selection of a reference voltage ($R_1<R_2<R_3$), each sub-SA performs a neat voltage comparison with RBL voltage and generates (N)OR3, (MAJ)MIN, and (N)AND3 logic functions simultaneously. The XOR-based comparison is then achieved through an interesting observation in which the three input majority function of OR3, MIN, and AND3 is able to generate XOR3 logic. The Boolean logic of in-memory XOR3 can be given as $XOR3 (/Sum) = MAJ((A+B+C) + (\overline{AB+AC+BC}) + (ABC))$. This unit is implemented with a low overhead capacitive voltage divider as shown in Fig. \ref{arc}(g). The implementation of 2-input bit-wise operations is straightforward by initializing one row to ``0''/``1''. 
We choose an 8T SRAM cell as a fast and compact design considering that the proposed in-memory computing mechanism operates based on BL discharging. Nevertheless, the mechanism presented here can be applied to various read-write decouple SRAM designs.

From a programmer's perspective, NS-LBP is more of a third party accelerator that can be connected directly to the memory bus or through PCI-Express lanes rather than a memory unit. Therefore, a virtual machine and ISA for general-purpose parallel thread execution need to be defined. We designed instruction sets that could optimally leverage highly parallel NS-LBP's operations discussed and developed a compilation framework on top of that. Accordingly, the programs will be translated at install time to the NS-LBP's hardware ISA tabulated in Table \ref{isa}. 

\begin{table}[t]
\centering 
\caption{NS-LBP ISA.}\vspace{-1em}
\scalebox{0.65}{
\begin{tabular}{|ccc|c|ccc|}
\hline
\multicolumn{1}{|c|}{Opcode}                 & \multicolumn{1}{c|}{Src1} & Src2 & Src3                                                                    & \multicolumn{1}{c|}{Dest} & \multicolumn{1}{c|}{Size} & Description                                  \\ \hline
\multicolumn{1}{|c|}{{\tt NS-LBP copy}}         & \multicolumn{1}{c|}{r1}   & -    & -                                                                       & \multicolumn{1}{c|}{r2}   & \multicolumn{1}{c|}{n}    & r2[i] = r1[i]                                \\ \hline
\multicolumn{1}{|c|}{{\tt NS-LBP ini}}          & \multicolumn{1}{c|}{r1}   & -    & \begin{tabular}[c]{@{}c@{}}a[i] = all '0'\\ b[i] = all '1'\end{tabular} & \multicolumn{1}{c|}{-}    & \multicolumn{1}{c|}{n}    & r1[i] = a[i]  or  r1[i] = b[i]               \\ \hline
\multicolumn{1}{|c|}{{\tt NS-LBP cmp (xor2)}}   & \multicolumn{1}{c|}{r1}   & r2   & a[i] = all '0'                                                          & \multicolumn{1}{c|}{r3}   & \multicolumn{1}{c|}{n}    & r3[i] = r1[i] $\oplus$ r2[i]                 \\ \hline
\multicolumn{1}{|c|}{{\tt NS-LBP search}}       & \multicolumn{1}{c|}{r1}   & k    & -                                                                       & \multicolumn{1}{c|}{r3}   & \multicolumn{1}{c|}{n}    & r3[i]= (r1[i] == k)                          \\ \hline
\multicolumn{1}{|c|}{{\tt NS-LBP nand3}}        & \multicolumn{1}{c|}{r1}   & r2   & r3                                                                      & \multicolumn{1}{c|}{r4}   & \multicolumn{1}{c|}{n}    & r4[i]= !(r1[i] \& r2[i]  \& r3[i])           \\ \hline
\multicolumn{1}{|c|}{{\tt NS-LBP nor3}}         & \multicolumn{1}{c|}{r1}   & r2   & r3                                                                      & \multicolumn{1}{c|}{r4}   & \multicolumn{1}{c|}{n}    & r4[i] = !(r1[i] $\|$ r2[i] $\|$ r3[i])         \\ \hline
\multicolumn{1}{|c|}{{\tt NS-LBP carry (maj3)}} & \multicolumn{1}{c|}{r1}   & r2   & r3                                                                      & \multicolumn{1}{c|}{r4}   & \multicolumn{1}{c|}{n}    & r4[i] = maj(r1[i], r2[i], r3[i])             \\ \hline
\multicolumn{1}{|c|}{{\tt NS-LBP sum (xor3)}}   & \multicolumn{1}{c|}{r1}   & r2   & r3                                                                      & \multicolumn{1}{c|}{r4}   & \multicolumn{1}{c|}{n}    & r4[i] = r1[i] $\oplus$  r2[i] $\oplus$ r3[i] \\ \hline
\multicolumn{3}{|c|}{r1-r4: addresses}                                          & k:address                                                               & \multicolumn{3}{c|}{$\forall i, i \in [1, n] , X = [ 64/128/256 ]$}                                    \\ \hline
\end{tabular}}
\label{isa}\vspace{-2em}
\end{table}

\vspace{-1em}
\subsection{In-memory LBP Algorithm}
\label{algo}
By converting a conventional software-based sequential comparison operation into a parallel bit-wise XOR operation, we propose an NS-LBP hardware-oriented LBP algorithm that fully utilizes the sub-array parallelism of the NS-LBP. A key objective in developing such an algorithm is to enable a parallel bit-position-aware comparison between pivot (C) and surrounding pixels (P) and generate an LBP bit-stream in fewer cycles, eliminating unnecessary power-hungry bulk bit-wise operations. For every LBP kernel, starting from the Most Significant Bit (MSB), Algorithm 1 issues the NS-LBP's comparison command (NS-LBP\_XOR) in a loop to pivot and pixels in parallel and update the Result\_array (line-7). The result of $i^{th}$ bit comparison ($C_i \oplus P_{j,i}$) is leveraged as a determining factor for NS-LBP to take the next step. As indicated in the algorithm, when the XOR result is ``1'', i.e., two unequal bits are identified (line-8), $C_i$ is read (NS-LBP\_Mem). Now, if $C_i$ equals ``0'', the corresponding LBP\_array position is set by ``1'', indicating $C_i<P_{j}$ and vice versa (lines-9-12). However, if equality is noticed, the next less significant bit in pixels and pivot is selected for comparison, and this process stops when the XOR result is ``1'' (inequality). Such a parallel comparison operation could rapidly detect the mismatch between all pixels and pivot from MSB to LSB. Our algorithm has a constant search time that is determined by the bit length of numbers. As shown, NS-LBP\_XOR is iteratively used in a nested ``for'' loop in the algorithm, and the NS-LBP architecture is mainly designed to accelerate this operation.  \vspace{-0.5em}

\begin{figure}[h]
\begin{center}
\begin{tabular}{c}
\includegraphics [width=0.97\linewidth]{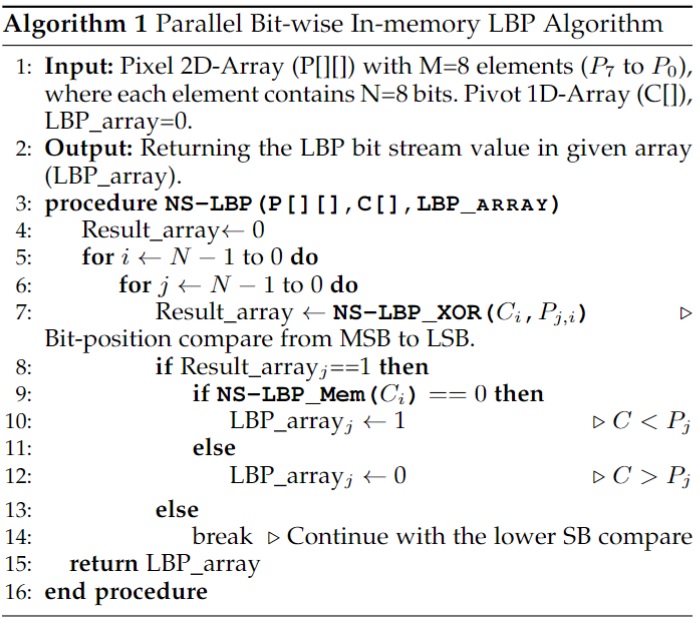} \vspace{-0.7em}\\
 \end{tabular} 
\vspace{-1.2em}
\end{center}\vspace{-0.5em}
\end{figure}


\section{Correlated Hardware Mapping} \vspace{-0.5em}
\subsection{LBP Layer} 
\begin{figure}[t]
\begin{center}
\begin{tabular}{c}
\includegraphics [width=0.99\linewidth]{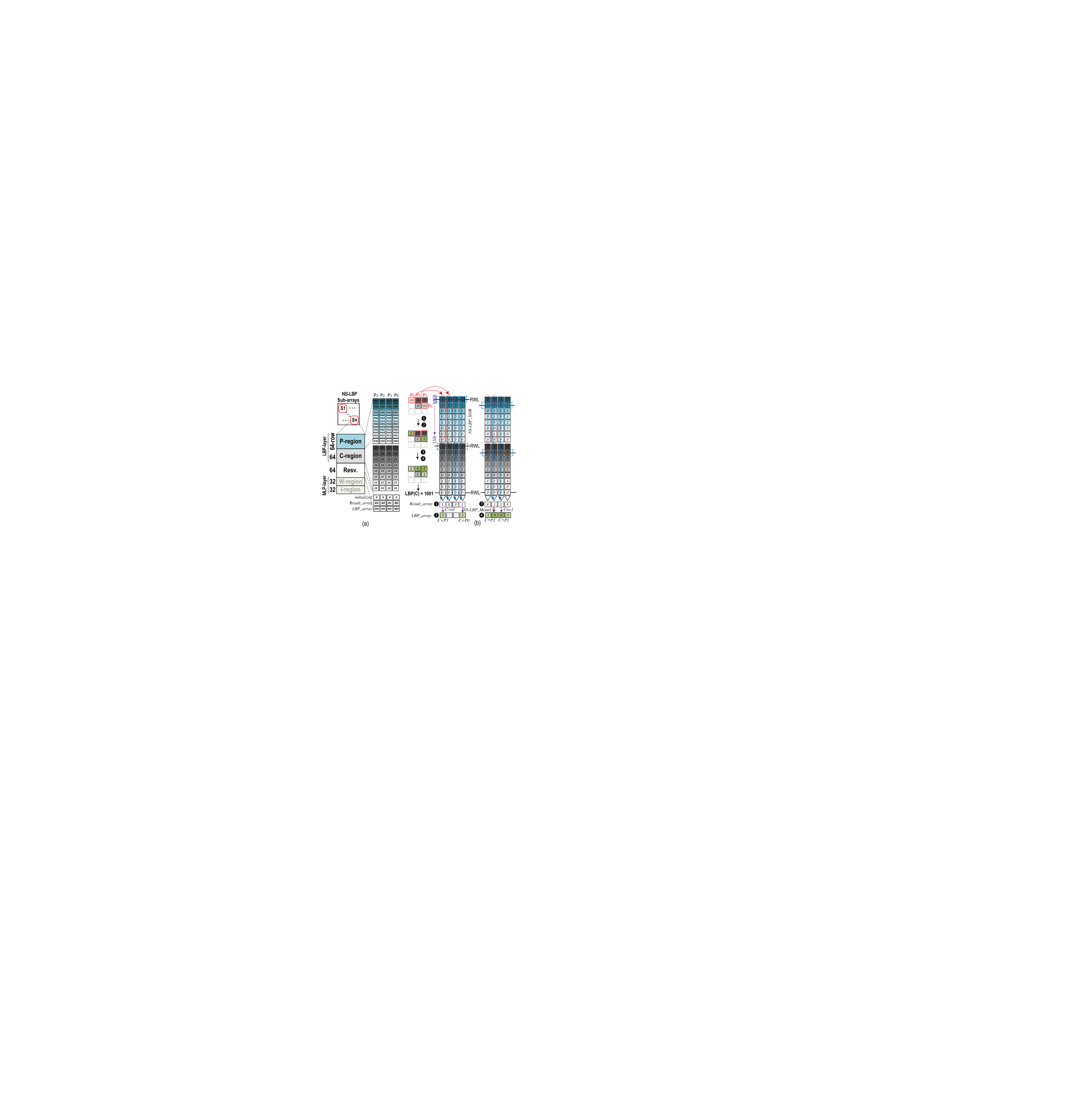} \vspace{-0.7em}\\
 \end{tabular} 
 \caption{(a) The NS-LBP's correlated data partitioning and mapping scheme, (b) Parallel LBP computation in NS-LBP. } \vspace{-1.2em}
\label{map}
\end{center}\vspace{-0.5em}
\end{figure}

\vspace{-0.5em}
To maximize Ap-LBP computation throughput and fully leverage NS-LBP's parallelism, we propose partitioning data as shown in Fig. \ref{map}. Given an LBP layer, the accessed memory region of pixels and pivots could be easily predicted, and the LBP bit-stream could be locally computed if we could store such correlated regions into the same memory sub-array.  
Thus, we propose a novel, correlated data partitioning, and mapping methodology as shown in Fig. \ref{map}(a) to locally store correlated regions of pixel and pivot vectors in the same memory sub-array and enable entirely local computation (i.e., NS-LBP\_XOR and NS-LBP\_Mem completely within the same sub-array without inter-bank/chip communication). 
The NS-LBP's compute sub-array (256 rows$\times$256 columns) is split into five key regions, i.e., Pixel-P (64 rows), Pivot-C (64 rows), Reserved (64 rows), Weight-W (32 rows), and Input-I (32 rows). We use P-, C-, and Resv. regions to process the LBP layer. 

The selected input pixels in Ap-LBP are initially transposed in the NS-LBP's buffer and mapped into P-region. In addition, we propose to store $P_{i+1}$ transposed copy of pivot as reference vectors in C-region. The C-region is specially designed to enable fully parallel bit-wise position-aware comparison operation.
Three rows in the Resv. region are dedicated to Result\_array, LBP\_array, and all-zero.  
Figure \ref{map}(b) gives an intuitive example of LBP-layer computation with the in-memory LBP algorithm, where four pixels ($P3$ to $P_0$) are selected. After data mapping, NS-LBP's Ctrl activates three RWLs simultaneously, corresponding to pixels' and pivot's MSB and all-zero row. 
The NS-LBP sub-array then performs the parallel XOR2 operation in a single cycle based on the mechanism discussed in Section 4, and the result ``1001'' is stored in the Result\_array row (step \encircle{1}). Now, the Ctrl readily recognizes the potential mismatch in $P_3$ and $P_0$ and accordingly updates LBP\_array in step \encircle{2} (``1xx1'') with respect to $C_7$=0 value. As there are two matches ($P_{2,7}$-$C_{7}$ and $P_{1,7}$-$C_{7}$), the Ctrl selects the next MSBs in pixels and pivot to find the next potential mismatch. The final LBP\_array value  (``1001'') is returned in step \encircle{4} for the next step.

\subsection{MLP Layer}
\begin{figure}[t]
\begin{center}
\begin{tabular}{l}
\includegraphics [width=0.96\linewidth, height=4.8cm]{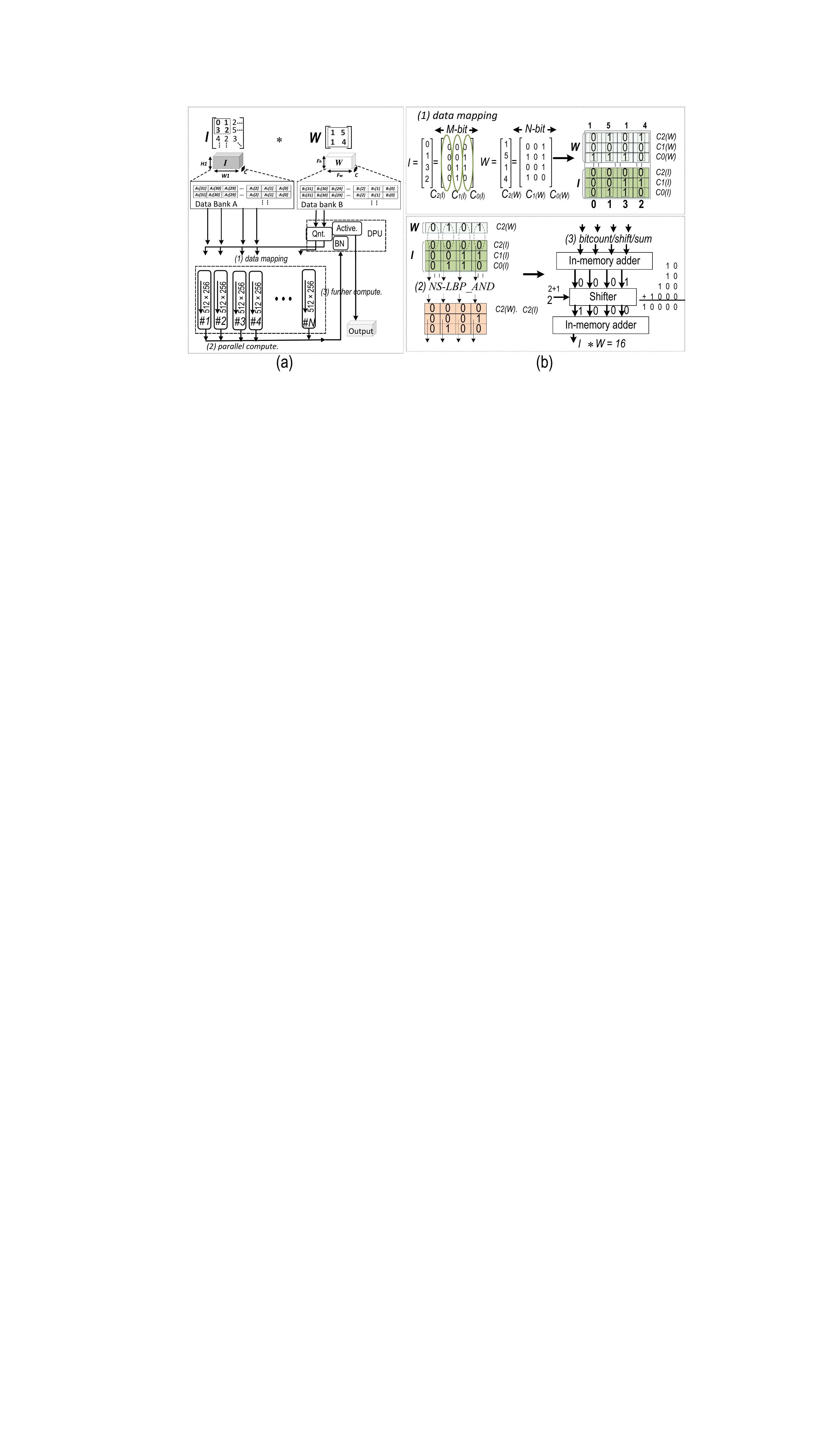} \\
 \end{tabular}
\caption{Parallel MLP computation in NS-LBP.} \vspace{-0.2em}
\label{PIM}
\end{center}
\end{figure}

Besides the LBP layer, there are MLP layers in Ap-LBP as shown in Fig. \ref{lbp}(b) that can be accelerated close to the sensor without sending the activated LBP feature maps to an off-chip processor. Note that MLP can be equivalently implemented by convolution operations using $1\times1$ kernels \cite{zhou2016dorefa}.
W- and I- regions in every NS-LBP sub-array (Fig. \ref{map}(a)) are dedicated to performing such an operation locally.
Figure \ref{PIM} gives an overview of the MLP bit-wise acceleration steps.
In the first step, the processed input activation from NS-LBP's LBP layers is quantized by DPU and mapped into I-region, where the MLP layer weights are located. In the second step, parallel computational sub-arrays perform bulk bit-wise operations between tensors and generate the output. Then, the output is activated by DPU's Activation unit and saved back into the Resv. region.
From a computation perspective, every MLP layer can be equivalently implemented by exploiting NS-LBP\_AND, bitcount, and bitshift as parallelizable operations \cite{zhou2016dorefa}.
Assume $I$ is a sequence of $M$-bit input integers, e.g., 3-bit in Fig. \ref{PIM} located in ifmap covered by a sliding kernel of $W$, such that $I_i \in I$ is an $M$-bit vector representing a fixed-point integer. 
We index the bits of each $I_i$ element from LSB to MSB with $m=[0 , M-1]$, such that $m=0$ and $m=M-1$ are corresponding to LSB and MSB, respectively. Accordingly, we represent a second sequence denoted as $C_m(I)$ including the combination of $m^{th}$ bit of all $I_i$ elements (shown by colored elliptic). For instance, $C_0(I)$ vector consists of LSBs of all $I_i$ elements ``0110''. 
Considering $W$ as a sequence of $N$-bit weight integers (3-bit, herein) located in a sliding kernel with an index of $n=[0, N-1]$. The second sequence can be similarly generated as $C_n(W)$. Considering the set of all $m^{th}$ value sequences, the $I$ can be represented like $I=\sum_{m=0}^{M-1}2^mc_m(I)$. Likewise, $W$ is represented like $W=\sum_{n=0}^{N-1}2^nc_n(W)$. Thus, the convolution between $I$ and $W$ is defined as $\small \sum _{m=0}^{M-1}\sum _{n=0}^{N-1}2^{m+n}bitcount(and(C_n(W),C_m(I)))$ \cite{zhou2016dorefa}. 
In the data mapping step of Fig. \ref{PIM}, $C_2(W)$-$C_0(W)$ and $C_2(I)-C_0(I)$ are consequently mapped into an NS-LBP's sub-array. Now, a parallel bit-wise AND operation ({NS-LBP\_AND}) of $C_n(W)$ and $C_m(I)$ is performed. The results will be then processed using a bit-counter counting the number of ``1''s in each vector and then a shifter unit, e.g., here left-shifted by 3-bit ($\times2^{2+1}$) to ``1000''. Eventually, the shifter unit's outputs are added up to produce ofmaps for every layer.\vspace{-1.2em}

\section{Evaluation Results}
\subsection{Setup}
\begin{figure}[t]
\begin{center}
\begin{tabular}{l}
\includegraphics [width=0.98\linewidth, height=4.5cm]{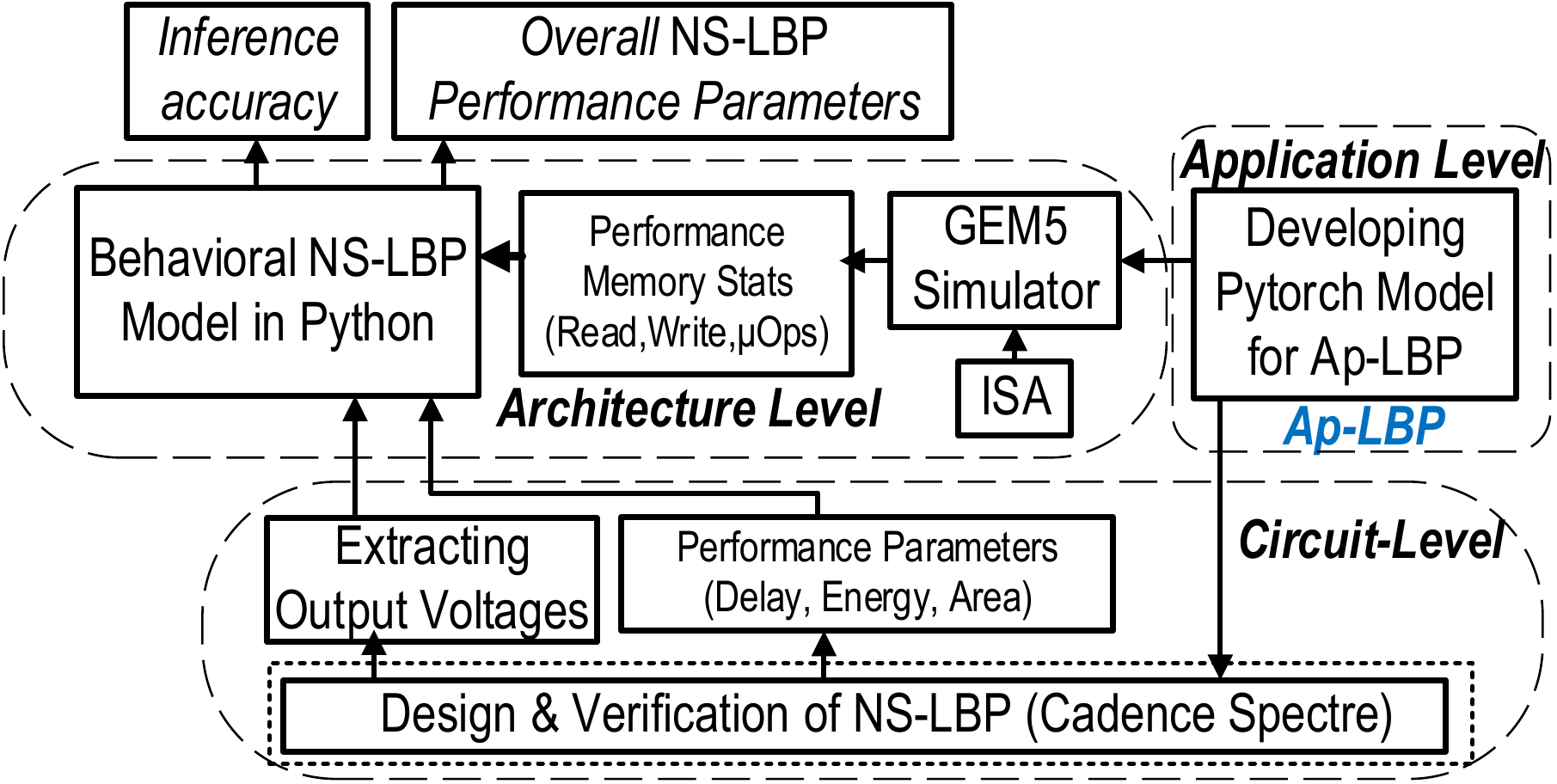} \vspace{-0.2em}\\
 \end{tabular}\vspace{-0.4em}
\caption{Evaluation framework developed for NS-LBP accelerator.}
\label{Evalf}\vspace{-2em}
\end{center}
\end{figure}

To estimate the performance of NS-LBP along with Ap-LBP, a bottom-up evaluation framework is developed as shown in Fig. \ref{Evalf}. At the \textit{circuit level}, NS-LBP is fully implemented with TSMC 65nm-GP with a supply voltage of 0.9V-1.1V in Cadence, and the post-layout results are reported. However, NS-LBP is not yet taped out.
At the \textit{architecture level}, we fully implemented NS-LBP's ISA using gem5 \cite{binkert2011gem5} and export the memory statistics and performance into a behavioral NS-LBP's in-house optimizer tool, also taking the circuit-level data to model the timing, energy, and area. This tool will offer the same flexibility in memory configuration regarding bank/mat/sub-array
organization and peripheral circuitry design as Cacti \cite{thoziyoor2008cacti} while supporting SRAM-level configurations. The architecture simulator can alter the configuration files with different array organizations. At the \textit{application level}, we trained a PyTorch implementation of Ap-LBP inspired by LBPNet, with the difference that our design approximates pre-trained LBP kernel parameters. The Ap-LBP's statistics are then leveraged in the behavioral NS-LBP model to compute the latency and energy of the whole system. Besides, to model the data loading time for all layers, we followed the approach in \cite{eckert2018neural} by developing a micro-benchmark that sequentially accesses the sets in a way that requires data loading. 

\begin{figure}[b]
\begin{center}
\begin{tabular}{c}
\includegraphics [width=0.99\linewidth]{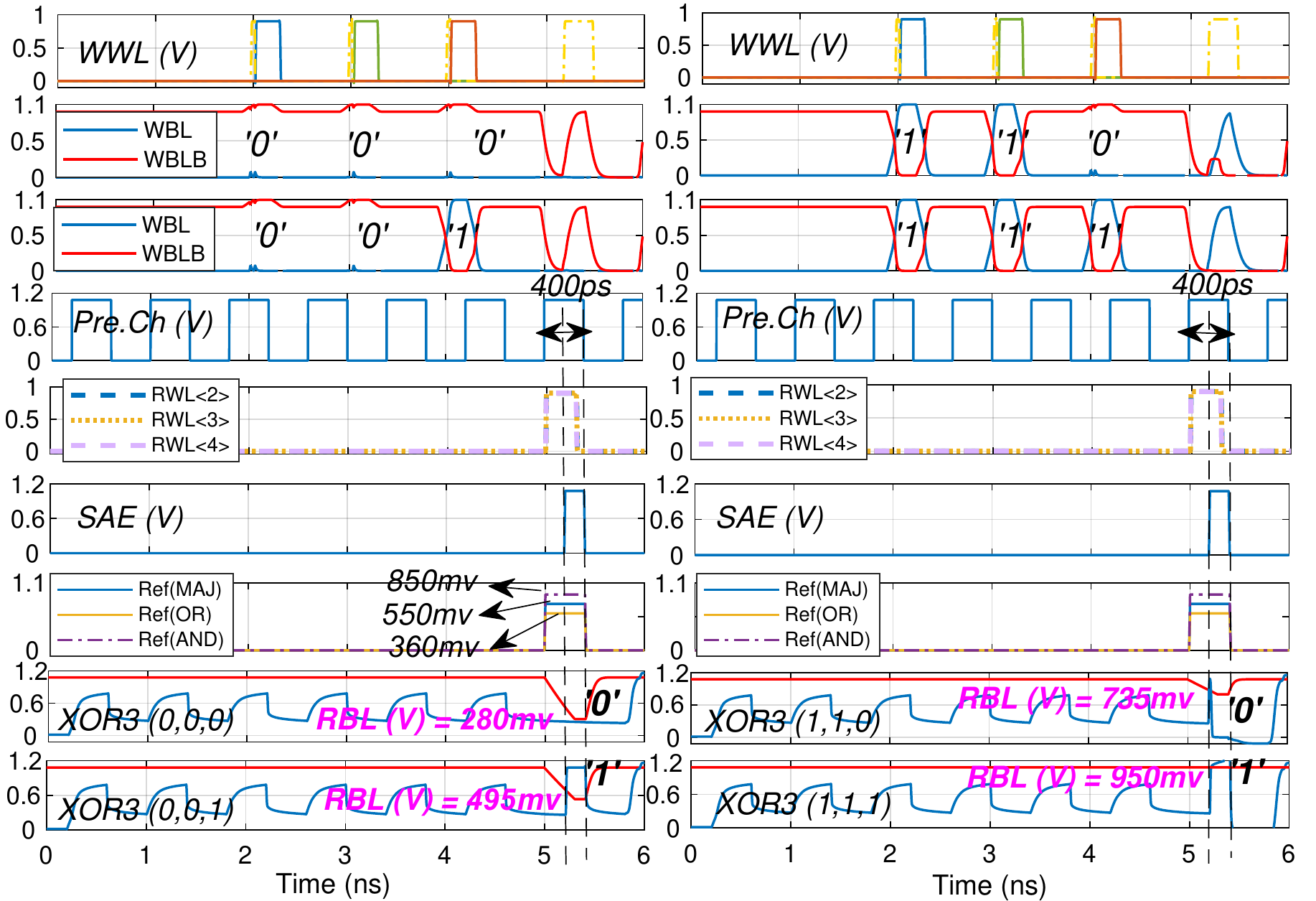}\\
 \end{tabular}
\caption{Transient simulation results of an NS-LBP sub-array executing comparison (based on XOR3) operation. }
\label{wave}
\end{center}
\end{figure}

\subsection{Functionality Analysis}
Figure \ref{wave} shows the post-layout transient simulation results of an NS-LBP sub-array. To verify the functionality of all possible input combinations (``000'', ``001'', ``011'', and ``111''), three WWLs are activated consecutively (first waveform) and by assigning proper voltages to WBL and WBLB, the SRAM cells are loaded with the operands. In the computation mode, we simultaneously activate the corresponding RWL of three cells to discharge RBL from the precharged voltage (1.1V) w.r.t. the memory value. To compromise three-row activation stability by lowering the RWL voltage that leads to read latency, we reduced the RWL voltage to 790mV to achieve the industry standard 6-sigma margin. 
For the evaluation, by activating the Sense Amplifier Enable (SAE) signal, a voltage comparison between the RBL voltage and references is made. As shown in Fig. \ref{wave}, $V_{R1}$=360mV, $V_{R2}$=550mV, and $V_{R3}$=850mV are set as the reference voltages.

In the case of ``000'', T8s (see Fig. \ref{arc}(d)) of all three cells are ON pulling down the RBL voltage from 1.1V to 280mV. This can be easily detected by SA generating ``0'' as the XOR3 output ($V_{R3}>V_{R2}>V_{R1}>$280mV). The total processing time from enabling the SA to get the result is $\sim$400ps in the same range given by the standard foundry memory compiler. In the case of ``001'', T8s of two out of three cells are ON pulling down the RBL voltage from 1.1V to 495mV. This can be easily detected by SA generating ``1'' as the XOR3 output ($V_{R3}>V_{R2}>$495mV$>V_{R1}$). With ``011'', T8 of only one cell is ON pulling down the RBL voltage from 1.1V to 735mV generating ``0'' as the XOR3 output ($V_{R3}>$735mV$>V_{R2}>V_{R1}$). Eventually, with ``111'' as inputs, all T8s are OFF taking the RBL voltage at 950mV outputting ``1'' (950mV$>V_{R3}>V_{R2}>V_{R1}$). 

\begin{table*}[t]
\begin{center}
\caption{Comparison with previous processing-in-SRAM accelerators.}\vspace{-1.2em}
\scalebox{0.75}{
\begin{tabular}{cccccccc}
\hline
Reference                & NS-LBP                     & Symp. VLSI \cite{valavi2018mixed} & DAC’20 \cite{lee2020bit} & JSCC’20 \cite{jiang2020c3sram} & JSSC’19 \cite{wang201928} & DAC’19 \cite{simon2019fast} & ISSCC’19 \cite{yang201924}                                                 \\ \hline
Technology               & 65nm                     & 65nm                             & 28nm                     & 65nm                           & 28nm                     & 28nm                        & 28nm                                                                       \\
Bitcell  Density         & 8T                       & 10T1C                            & 6T                       & 8T-1C                          & 8T Transposable          & 6T/local group              & 8T                                                                         \\
SA compute Area Overhead & 3.4$\times$              & -                                & 4.94$\times$             & -                              & 5.52$\times$             & 5.05$\times$                & $>$15$\times$                                                                \\
LBP-comparison Support   & Yes                      & No                               & No                       & No                             & Yes                       & Yes                         & No                                                                         \\
MAC Support              & Yes (digital CNN)        & Yes (analog BWNN)                & Yes (digital CNN)        & Yes (analog BWNN)              & Yes (digital CNN)        & No                          & Yes (analog BWNN)                                                          \\
Supply                   & 0.9V-1.1V                & 0.68-1.2V                        & 0.6V-1.1V                & 0.6V-1V                        & 0.6V–1.1V                & 0.6V –1.1V                  & 0.6-0.9V                                                                   \\
Max Frequency            & 1.25GHz (1.1V)             & 100MHz                           & 2.25GHz (1V)             & 50MHz                          & 475MHz (1.1V)            & 2.2GHz (1.0V)               & 400MHz                                                                     \\
TOPS/W                   & 37.4                      & 658                              & 8.09 (0.6V, 372MHz)      & 671.5                          & 5.27 (0.6V, 114MHz)      & -                           & 5.83                                                                       \\
Array size               & 4$\times$256$\times$ 256 & -                                & 4$\times$128$\times$ 128 & 4$\times$128$\times$ 128       & 4$\times$128$\times$256  & 256$\times$64               & \begin{tabular}[c]{@{}c@{}}4*I: 28x28x4x6x8/\\ 4*W: 28x28x4x2\end{tabular} \\ \hline
\end{tabular}}\vspace{-2.2em}
\label{inSRAM}
\end{center}
\end{table*}

\begin{figure}[t]
\begin{flushleft}
\begin{tabular}{c}
\includegraphics [width=0.95\linewidth]{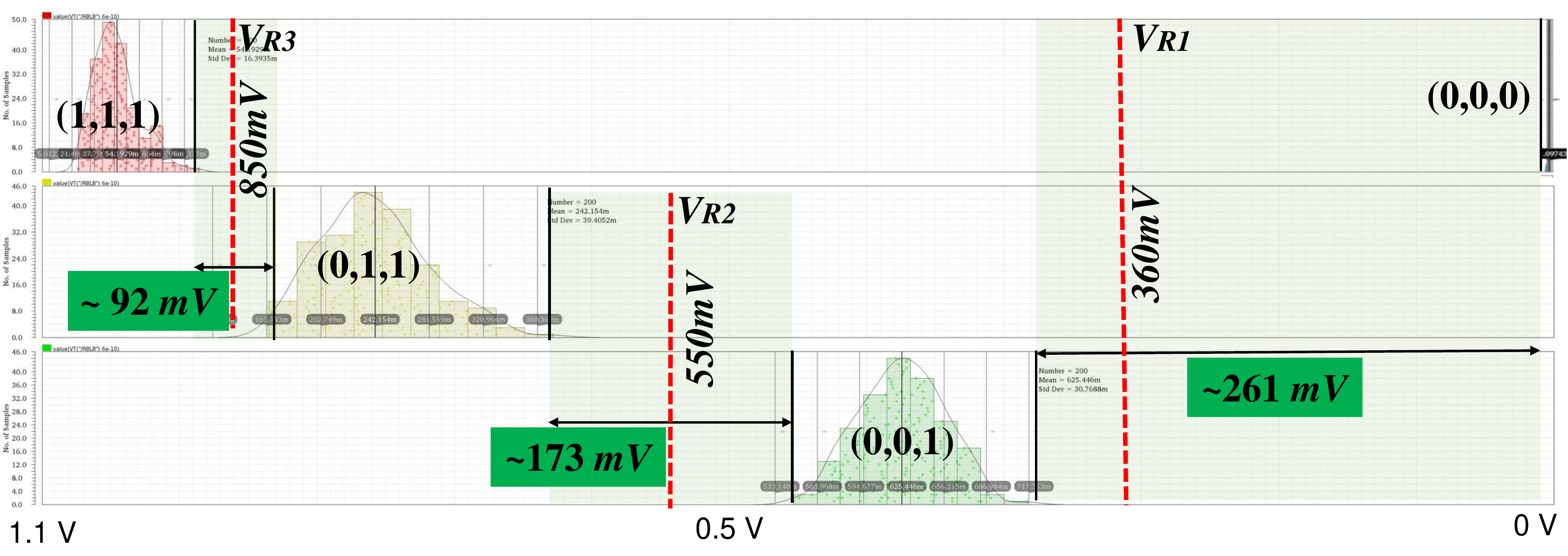}
\end{tabular}\vspace{-0.9em}
\caption{Monte-Carlo simulation of RBL and SA reference voltage.}
\label{MC_Sim}\vspace{-1.2em}
\end{flushleft}
\end{figure}

For the SA reference voltage ($V_{R}$) analysis, the RBL sense margins are first tested through post-layout Monte Carlo simulations in Cadence Spectre, as shown in Fig. \ref{MC_Sim}, where the sensing margin is reported considering both process (inter-die) and mismatch variations (intra-die) for core VDD (1.1 V) at 1.25 GHz. To conduct the $V_{R}$ variation analysis, we tested all 256 bit-lines within each NS-LBP's sub-array, 200 times, for all possible bit value combinations in memory. It is found that at lower voltages the maximum operating frequency is limited by the reduction of $V_{Ref}$ ranges. A higher VDD also yields a larger sensing margin. As we observe there is $\sim$92mV margin (the smallest voltage margin observed between ``111'' and ``011'' cases) between each two combination bringing a high in-memory computing reliability for the NS-LBP design. 

\begin{figure}[b]
\begin{center}
\begin{tabular}{l}
\includegraphics [width=0.97\linewidth]{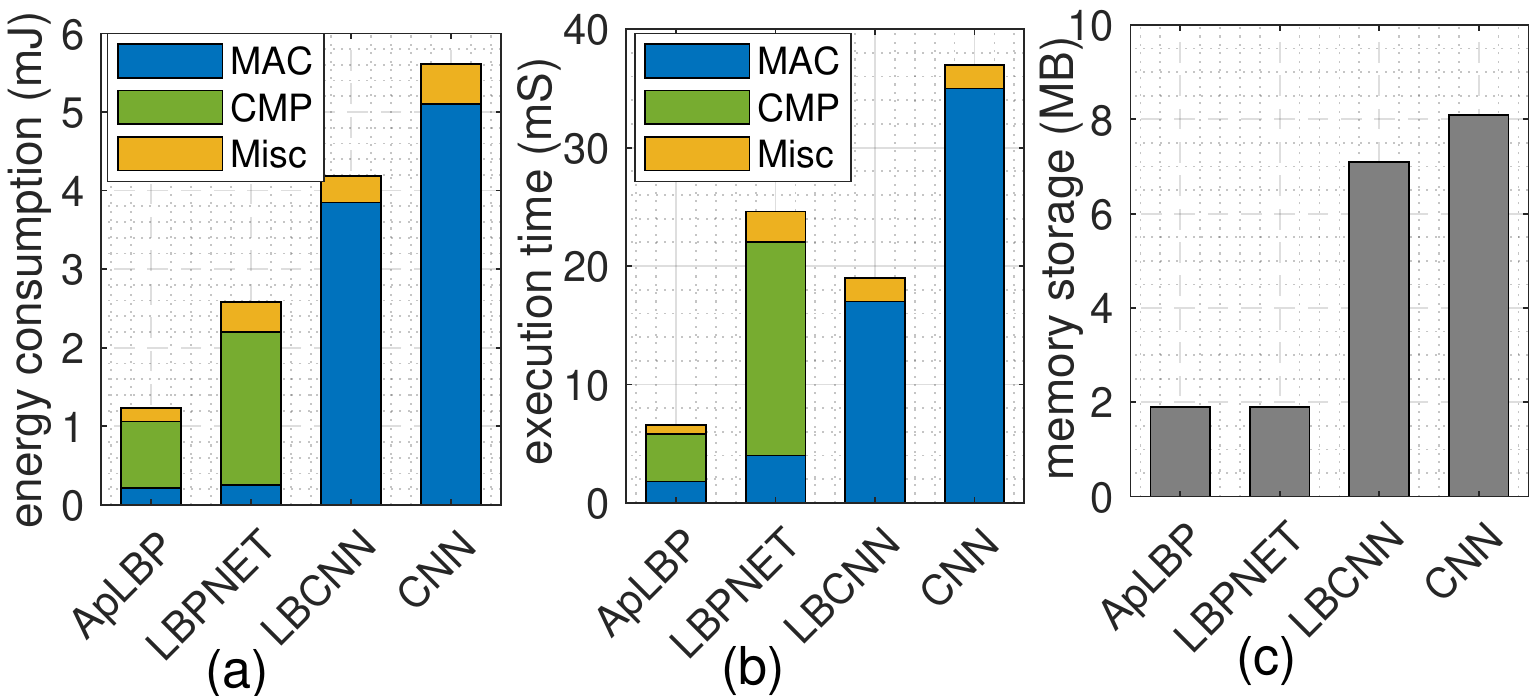} \\
 \end{tabular}\vspace{-1.2em}
\caption{(a) Energy consumption, (b) Execution time, and (c) Memory storage comparison.}
\label{pow}\vspace{-1.2em}
\end{center}
\end{figure}

\subsection{Energy Consumption \& Performance} Figure \ref{pow}(a) shows the energy consumption breakdown of NS-LBP running Ap-LBP and LBPNet compared to a baseline 8-bit quantized CNN and LBCNN implemented by \cite{wang201928} running SVHN data-set. We meticulously report the energy consumed by MAC and CMP operations in various networks.
We observe that (i) the NS-LBP running Ap-LBP demonstrates up to $\sim$2.2$\times$ and 5.2$\times$ better energy efficiency compared to the LBPNet and CNN counterparts, respectively. Converting power-hungry MAC to bit-wise comparison operation in an approximate fashion has yielded such a striking improvement; (ii) leveraging Ap-LBP can bring up to $\sim$4$\times$ energy-efficiency when compared with the LBCNN. It is worth mentioning that LBCNN still relies on power-hungry MAC operations.
Figure \ref{pow}(b) compares inference delay per input image in four under-test designs. We observe that the NS-LBP leveraging Ap-LBP achieves $\sim$4$\times$ and 2.3$\times$ speed-up compared to LBPNet and LBCNN designs, respectively. Besides, it can be seen that $\sim$6.2$\times$ speed-up is achieved when compared with the CNN baseline. 
Figure \ref{pow}(c) further clarifies that Ap-LBP doesn't remarkably reduce the memory storage relative to LBP-Net; however, it requires $\sim$3.4$\times$ smaller memory to store the parameters than LBCNN. \vspace{-0.1em}

\vspace{-0.8em}
\subsection{Comparison} Since several processing-in-SRAM platforms have been developed to accelerate various deep neural networks in literature, performing a fair comparison is almost difficult. Nevertheless, Table \ref{inSRAM} lists six recent designs for comparison. As can be seen, various designs are implemented with different bit-cell structures and SA designs. Here we report our main observations. 
(1) The NS-LBP and the designs in \cite{simon2019fast,wang201928} are the only in-SRAM platforms that can support XOR-based LBP computation. We observe that NS-LBP shows a fairly smaller SA area overhead (3.4$\times$) compared to these designs to support in-memory computation. 
(2) It can be seen that the designs presented in \cite{lee2020bit,simon2019fast} show the highest frequency at 1V, where NS-LBP stands as the third-fastest design. (3) The NS-LBP achieves 37.4 TOPS/W standing as the third most efficient design compared to all counterparts, whereas the design in \cite{jiang2020c3sram} with 671 TOPS/W stands as the most efficient design.

Overall, NS-LBP offers
1) A dual-mode computational SRAM platform with no sacrifice of memory capacity that directly processes data within the memory array to eliminate off-chip data communication; 2) A complete set of Boolean operations (both 2- and 3-input), majority, and full adder in only one single memory cycle, demonstrating one of the most efficient PNS systems to date; 3) Highly parallel low-bit-width convolution operation; and 4) Light modification of existing memory cell to achieve low in-memory logic area overhead. \vspace{-0.5em}

\subsection{Accuracy}
To perform a fair comparison between Ap-LBP and five other neural network models, CNN (as a baseline) \cite{gavrikov2022cnn}, Binarized Neural Network (BNN) \cite{baseline2}, BinaryConnect \cite{courbariaux2015binaryconnect},  LBCNN \cite{LBCNN}, and LBPNet \cite{LBPNET}, with identical hyper-parameters such as number of basic blocks,  number of hidden neurons, etc. are selected. We conduct experiments on three data-sets, i.e., MNIST, FashionMNIST, and SVHN, to evaluate the performance of both algorithm accuracy and hardware implementation. 
PyTorch implementation of Ap-LBP inspired by LBPNet, with the difference that our design approximates pre-trained LBP kernel parameters, is developed. The number of basic blocks for MNIST and SVHN is set to 5 (3 LBP layers and 2 FC layers) and 10 (8 LBP layers and two FC layers) layers, respectively, with 512 hidden neurons. The simulation is performed with two GPUs (Nvidia RTX 3090) configurations.
The comparison of classification accuracy is summarized in Table \ref{accuracy}. We examined two Ap-LBP variations, $\text{Ap-LBP}^{~(1)}$ and $\text{Ap-LBP}^{~(2)}$ with one and two approximated bits, respectively. Based on the obtained results, the Ap-LBP shows a minor accuracy degradation compared to counterpart networks while providing significant energy-product-delay reduction as discussed earlier. As can be seen $\text{Ap-LBP}^{~(2)}$ achieves the accuracy of 90.3\% on the SVHN data-set, while the LBCNN and LBPNet show 94.50\% and 92.90\% accuracy, respectively.

\begin{table}[h]
\footnotesize
\centering
\caption{Inference accuracy (\%) of LBP networks vs. CNN.}\vspace{-1em}
\scalebox{0.99}{
\begin{tabular}{lccc}
\hline
\rowcolor[HTML]{C0C0C0} 
Model                                             & MNSIT & FashionMNIST & SVHN  \\ \hline
Baseline \cite{gavrikov2022cnn}                   & 99.48 & 94.44        & 95.21 \\ \hline
BNN \cite{baseline2}                              & 98.60 & 91.86        & 97.49 \\ \hline
BinaryConnect \cite{courbariaux2015binaryconnect} & 98.99 & -            & 97.85 \\ \hline
LBCNN \cite{LBCNN}                                & 99.51 & -            & 94.50 \\ \hline
LBPNet \cite{LBPNET}                              & 99.50 & 90.61        & 92.90 \\ \hline
\textbf{$\text{Ap-LBP}^{ (1)}$}                   & 99.21 & 89.99        & 91.67 \\ \hline
\textbf{$\text{Ap-LBP}^{ (2)}$}                   & 97.98 & 86.93        & 90.3  \\ \hline
\end{tabular} \vspace{-1em}}
\label{accuracy}\vspace{-1em}
\end{table}

 \vspace{-0.5em}

\section{Conclusion}
This paper first presented an approximate and multiply–accumulate-free deep neural network model named Ap-LBP for efficient feature extraction. We then developed a {c}omparator-based {n}ear-{s}ensor {p}rocessing local binary pattern accelerator (NS-LBP) and a parallel in-memory LBP algorithm to process images near the sensor based on the Ap-LBP. 
The results on MNIST and SVHN data-sets demonstrate minor accuracy degradation compared to baseline CNN and LBP-network models, while NS-LBP achieves 1.25-GHz and energy-efficiency of 37.4 TOPS/W. NS-LBP reduces energy consumption and execution time by a factor of 2.2$\times$ and 4$\times$ compared to a recent LBP-based network.

\bibliographystyle{IEEEtran}
\scriptsize
\bibliography{IEEEabrv,./ref}

\end{document}